\documentclass{article}
\usepackage{spconf,amsmath,graphicx}

\title{Multi-View Networks For Multi-Channel Audio Classification}
\twoauthors
  {Jonah Casebeer$^{\sharp}$, Zhepei Wang$^{\sharp}$ \thanks{$^{\sharp}$These two authors contributed equally}}
	{University of Illinois at Urbana-Champaign\\
	Department of Computer Science \\
	jonahmc2, zhepeiw2@illinois.edu}
  {Paris Smaragdis \thanks{Supported by NSF grant \#1453104}}
	{University of Illinois at Urbana-Champaign\\
	Department of Computer Science \\
	Adobe Research}
\begin{document}
\ninept
\maketitle
\begin{abstract}
In this paper we introduce the idea of multi-view networks for sound classification with multiple sensors. We show how one can build a multi-channel sound recognition model trained on a fixed number of channels, and deploy it to scenarios with arbitrary (and potentially dynamically changing) number of input channels and not observe degradation in performance. We demonstrate that at inference time you can safely provide this model all available channels as it can ignore noisy information and leverage new information better than standard baseline approaches. The model is evaluated in both an anechoic environment and in rooms generated by a room acoustics simulator. We demonstrate that this model can generalize to unseen numbers of channels as well as unseen room geometries.
\end{abstract}
\section{Introduction}
\label{sec:intro}
Sound classification and detection is becoming an increasingly relevant problem, and one in which we are seeing a lot of activity and progress in the last few years. In this paper we focus on the case where we have a lot of acoustic sensors, but we are not guaranteed that they all record a clean enough signal for the task at hand, neither that they are all recording at any times. For instance, consider the case of a few people in an office setting. Each person will likely have a cell phone with a couple of microphones, a laptop with a few more, and maybe some mic-enabled wearable devices; there might be a room microphone tethered to a conferencing system, or audio-enabled desktop computers in the room, perhaps a few hearing aids, and maybe some security microphones as well. In this situation we might want to perform audio sensing tasks, e.g. diarization, and although we have access to a large number of recordings in the room, not all will be of use. For instance, some cell phones might be in purses providing a non-informative signal, whereas others might be right next to the type of sound we want to detect. Our goal is to explore algorithms that will not be misled by channels with low-informational content, and to not be tethered to a fixed number of channels. In order to do so we introduce the concept of the \textit{multi-view network} (MVN) for the purpose of classification. The framework that we present here considers a simple classifier, but is easily amenable to more elaborate extensions in order to facilitate state-of-the-art classification and detection models, as long as they fall under the umbrella of a deep learning model.

Deep learning models for monaural or binaural audio classification have been explored in many settings. Several deep architectures have shown to be powerful tools for the tagging task \cite{hershey2017cnn, vu2016acoustic, choi2016automatic, xu2017convolutional}. These techniques have been expanded to work in multi-channel scenarios as well.

Multi-channel audio classification and detection models mirror their single channel counterparts. In the DCASE-2018 Task 5 Challenge \cite{Dekkers2017}, the top performers used beamforming and source separation front ends with Convolutional Neural Network (CNN) ensemble back ends in combination with various data augmentation techniques \cite{inoue2018domestic, Tanabe2018}. Recurrent Neural Networks (RNNs)  {\cite{rnn}} operating on spectral features have been explored by \cite{parascandolo2016recurrent, kim2017acoustic} and others. Recently, \cite{li2016neural, xiao2016deep} used RNNs in an end-to-end fashion to predict the filters of a beamformers whose output fed to another deep model. 

In all of this work however, when a deep audio classification model is trained to perform classification on, e.g., four channels it typically can't guarantee the ability to leverage information when more channels are provided or to function when some are missing. In contrast to neural-based methods, classic beamforming approaches can scale to an arbitrary number of input channels. However, with a fixed number of channels available, learning methods are typically superior to beamforming. Here we seek to remedy this by using network architectures that accept inputs of variable sizes, allowing us to train on a fixed number of channels and deploy on any other number of channels at inference time.
Our work extends the Multi-View Networks for denoising \cite{mvn}. The denoising model attempted to predict clean spectra from noisy recordings, our classification model presented here extends that idea to that of performing classification. Our results show that MVNs are fit for classification and that they can handle channel disturbances in a dynamic environment with simulated Room Impulse Responses several times larger than our Short-Time Fourier Transform(STFT) frames.

\section{Multi-View Networks}
\label{sec:mvn}
RNNs are a common starting point for single channel audio classification. They typically operate on a chosen short-time spectral representation and unroll across time to leverage the temporal dependencies between successive spectral frames. Due to an RNN's ability to process inputs of arbitrary length, training and testing sequences for these models are not required to have a fixed length. 
This ability allows models to be trained on short sequences and operated on longer test sequences. In this work we study audio classification in a multi-channel scenarios where the number of channels available for training might differ from the number of channels available at test time.

\begin{figure}
    \centering
    \includegraphics[scale=0.3]{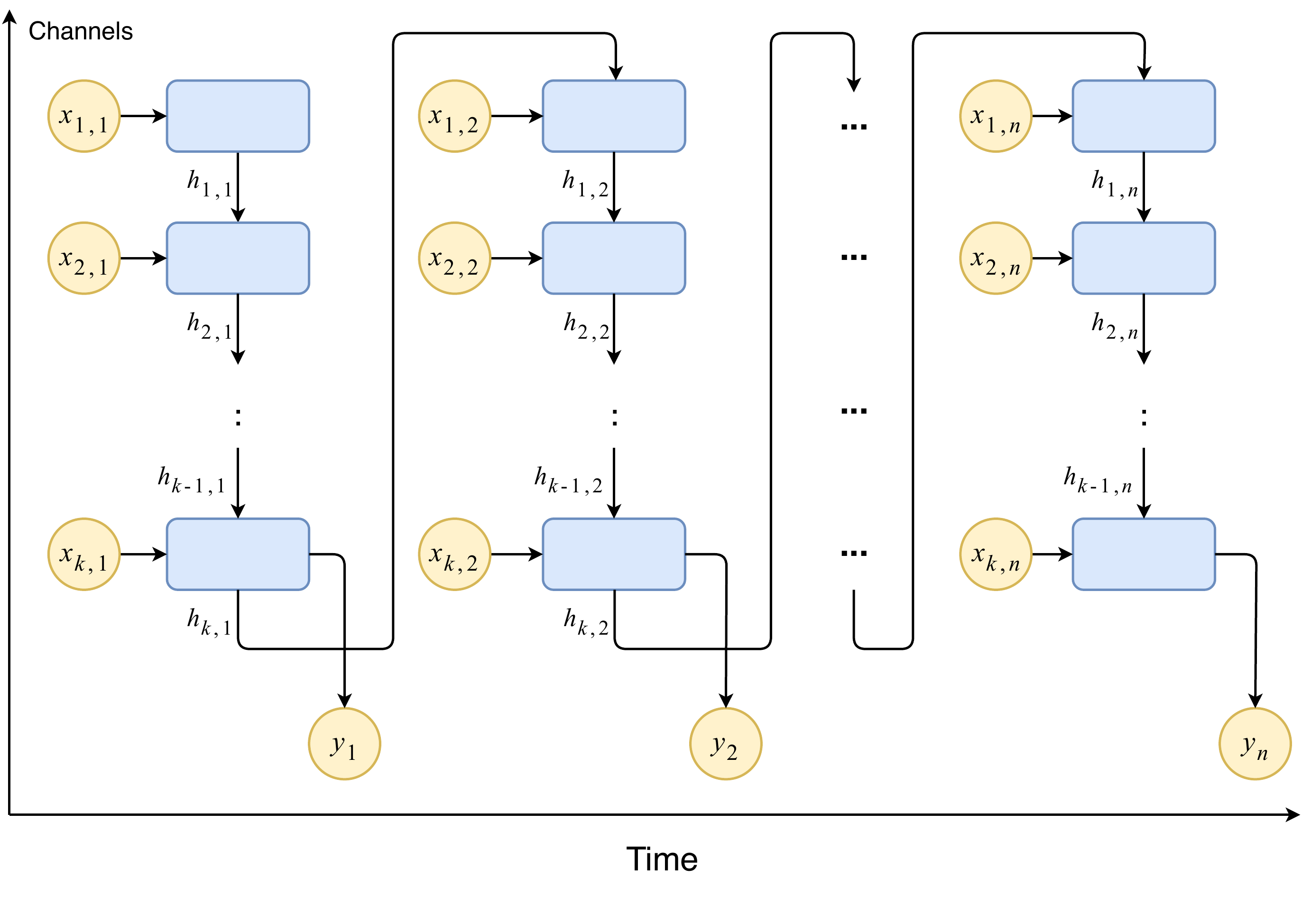}
    \caption{MVN unrolling across channels and time. Note that the model observes all shared time steps of each channel before making a prediction. Then, the last channel's hidden state feeds into the first channel of the next time step allowing for the propagation of temporal information.}
    \label{fig:mvn}
\end{figure}

Multi-View networks \cite{mvn} use the ability of RNNs to generalize to sequences of any length by unrolling across time and channels. RNNs perform the following operation using a non-linearity $\sigma$ where $\mathbf{x}_{k,t}$ is channel $k$'s $t$-th spectral frame:
\begin{align}
\begin{split}
    \mathbf{h}_t &=\sigma( \mathbf{W}_h \mathbf{x}_t +\mathbf{U}_h \mathbf{h}_{t - 1})\\
    \mathbf{y}_{t} &=\sigma(\mathbf{W}_x \mathbf{h}_{t})
    \end{split}
    \label{rnneq}
\end{align}
MVN's learn a set of matrices $\mathbf{W}_h, \mathbf{W}_x, \mathbf{U}_h$ to leverage temporal information and information across channels. This is achieved with the below recurrence.
\begin{align}
\begin{split}
    \mathbf{h}_{k, t} &= 
        \begin{cases}
            \sigma(\mathbf{W}_h \mathbf{x}_{k,t} + \mathbf{U}_h \mathbf{h}_{k,t-1}) & \text{if } k=1 \\
            \sigma(\mathbf{W}_h \mathbf{x}_{k,t} + \mathbf{U}_h \mathbf{h}_{k-1,t}) & \text{otherwise}
        \end{cases}\\
    \mathbf{y}_{t} &= \sigma(\mathbf{W}_x \mathbf{h}_{k,t})
\end{split}
\label{rnn2eq}
\end{align}
This recurrence allows the model to aggregate information from all channels at each time step before a prediction. Because this operation is fundamentally a single dimension RNN it can generalize to numbers of channels not seen before. Figure \ref{fig:mvn} demonstrates the unrolling across channels and time.
\section{Experiments}
\label{sec:experiments}
\label{ssec:disc}
\begin{figure}[htb]
\begin{minipage}[b]{1.0\linewidth}
  \centering
  \centerline{\includegraphics[width=8.5cm]{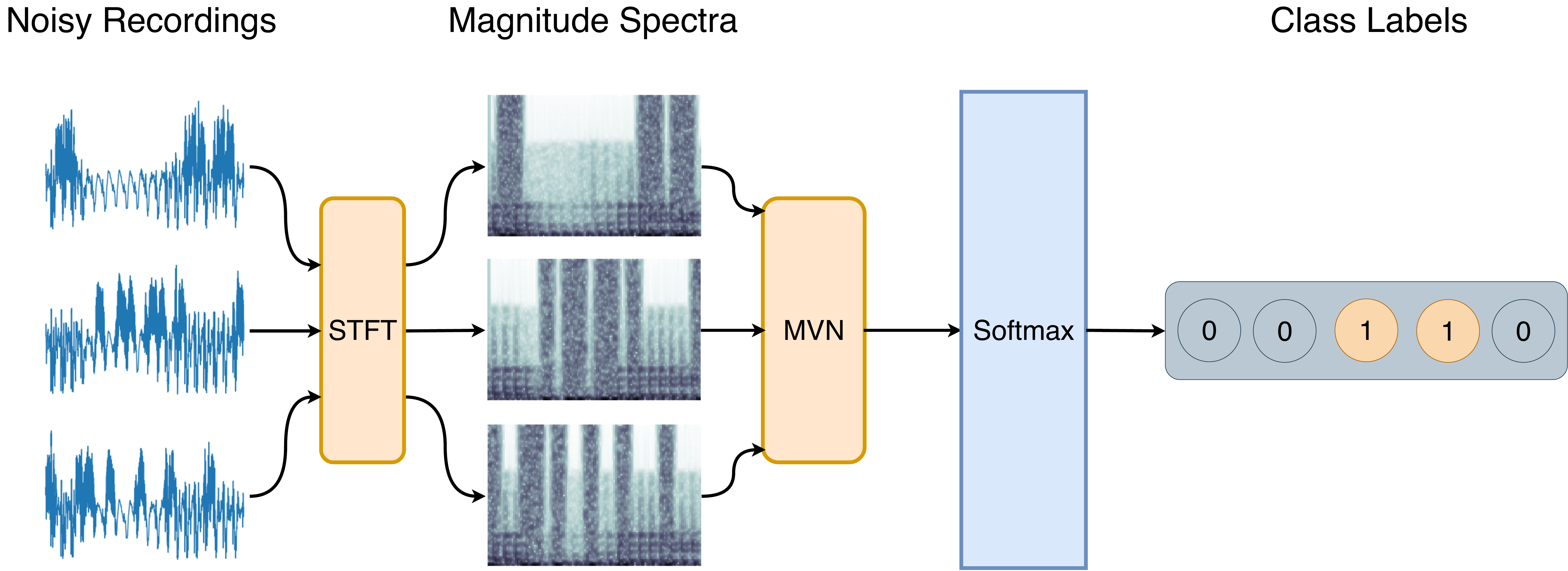}}
\end{minipage}
\caption{Audio event classification pipeline with an MVN. For a multi-channel mixture, we first take the magnitude spectra for each channel with STFT. The network takes the magnitude spectra as input and predicts if each frame contains speech.}
\label{fig:pipeline}
\end{figure}
In the following sections we introduce the setup of the audio classification experiments, including the data set, the baseline models, and two experiments evaluating the performance of the models. The models and the data correspond to a binary classification task in which each input is classified as either speech or not. Figure~\ref{fig:pipeline} illustrates the general pipeline for the experiment. These experiments model various scenarios with many microphones where most of the recorded signals are extremely noisy or contain little information. This reflects the potential application of this model in IoT sensing problems. 
\subsection{Data}
\label{ssec:dataset}
We prepare the data set for the binary audio classification experiments by mixing speech and noise segments. The speech segments are selected from the TIMIT Corpus, which includes 25,200 recordings with 630 speakers of 8 major dialects of American English, each reading 10 phonetically rich sentences \cite{timit}. The noise segments are selected from 13 different background noise recordings such as ``Airport'', ``Babble'' and ``Restaurant'' noises \cite{noise-16k}. 

Our data set is made up of two-second noisy mixtures at a sampling rate of 16kHz. To create a sequence, we randomly select a two-second segment from one of the noise recordings and a segment between zero to two seconds in length from one of the TIMIT speech recordings. Next, we normalize each of the noise and speech segments to unit variance. We then mix them by adding the speech segment to a random position within the two-second noise segment to obtain a single channel mixture. The resulting mixtures are approximately half speech and half background noise.

Based on this single channel mixture, we propose different ways to generate multi-channel mixtures; as described in Section~\ref{sssec:snr}. We apply STFT with a 1024 pt window and a 512 pt hop on each channel of the mixture, and take the absolute value of the coefficients as the input to the models.
\subsubsection{Generating Channels with Multiple SNR Values}
\label{sssec:snr}
In this experiment, we set the per channel SNR of a noisy mixture by scaling it to the desired SNR in decibels (dB). 
For the training and validation set, all mixtures contain four channels whose SNR values are evenly spaced between -5dB and 5dB.

For the test set, the number of channels is ranged between 2 and 30, emulating scenes with different numbers of available sensors in some ad-hoc network. We propose two scenarios for generating multi-channel mixtures for testing. In the first scenario, each additional channel has a lower SNR value than the prior channels. Specifically, for a mixture of $K \in \{2, 3, \ldots, 30\}$ channels, the SNR values for the channels are $0\text{dB}, -1\text{dB}, \ldots, -(K - 1)\text{dB}$. The more channels, the lower the average SNR value. We call this scenario ``decreasing SNR". We also propose an ``increasing SNR" scenario in which a mixture of $K \in \{2, 3, \ldots, 30\}$ channels contains SNR values of $-29\text{dB}, -28\text{dB}, \ldots, (-29 + K - 1)\text{dB}$. The more channels, the higher the average SNR value. In both scenarios, the channel indices are randomly shuffled.

By providing the model progressively worse channels we test the ability to ignore noisy information. By providing progressively better channels we test the ability to leverage new information. These setups expose the model to a diverse set of SNRS mimicking sensor networks in a chaotic environment. 
\subsection{Baseline Models}
\label{ssec:baseline}
We now introduce three different binary classification strategies as the baseline for the experiments. Each model takes the mixture's magnitude STFT spectra as input, and then pass it into a GRU with 512 hidden units unrolling across time followed by a softmax layer that classifies each input frame as either noise or signal.
\subsubsection{Averaging Input}
\label{sssec:avgin}
This model averages the magnitude spectra across channels for each mixture, and it then feeds the averaged spectra into the network. The output of the softmax layer is the estimated probability distribution of each frame being noise or signal:
\begin{equation}
    h_{\mathbf{\mathbf{\Theta}}}(\mathbf{X_t}) = \text{argmax}_{c \in \{0, 1\}} P(y_t=c | \overline{\mathbf{x_t}}; \mathbf{\Theta})
\end{equation}
\begin{equation}
    \overline{\mathbf{x_t}} = \frac{\sum_{k=1}^{K}\mathbf{x_{k,t}}}{K}
\end{equation}
where $\mathbf{X_t}$ is the set of magnitude spectra at time $t$ for a given mixture with $K$ channels, $\mathbf{x_{k,t}}$ is the spectra at time $t$ for the $k^{\text{th}}$ channel, $c \in \{0, 1\}$ is the binary label for each frame, and $\mathbf{\Theta}$ is the set of model parameters. We refer to this model as the Averaging Input model.
\subsubsection{Averaging Output}
\label{sssec:avgout}
This model takes the STFT coefficients for each channel of the mixture, and feeds it into the network. For a mixture with $K$ channels, we will therefore have $K$ different output probability distributions from the softmax layer. We obtain the probability for each frame by averaging the $K$ softmax probability distributions:
\begin{equation}
    h_{\mathbf{\Theta}}(\mathbf{X_t}) = \text{argmax}_{c \in \{0, 1\}}\frac{\sum_{k=1}^K P(y_t=c | \mathbf{x_{k,t}};\mathbf{\Theta})}{K}.
\end{equation}
We refer to this model as the Averaging Output model.
\subsubsection{Taking Output with Highest Probability}
\label{sssec:maxout}
Similar to the Averaging Output model, this model takes the STFT coefficients for each of the $K$ channels of the mixture and produces $K$ output distributions for each frame. Instead of averaging the output probabilities, we use the prediction with highest confidence: 
\begin{equation}
   h_{\mathbf{\Theta}}(\mathbf{X_t}) =  \text{argmax}_{c \in \{0, 1\}} \max_{k \in \{1 \ldots K\}} P(y_t = c|\mathbf{x_{k,t}}; \mathbf{\Theta}).
\end{equation}
%
We refer this model as the Max Output model.
\subsection{Training Setups}
\label{ssec:trnsetup}
For all experiments, we train with 250 batches of 40 k-channel mixtures per epoch(10k mixtures total). We use Adam \cite{adam} with an initial learning rate of $10^{-3}$ to minimize the cross-entropy loss, and we train each model for 100 epochs and use the model saved at the epoch with the lowest validation loss. The learning rate is decreased by a factor of 0.25 every 20 epochs. 
\subsection{Results}
 We now report the performance of the MVN and the baseline models in two experiments.
\subsubsection{Simple Mixtures}
\label{sssec:syn}
In the first experiment, we create multi-channel mixtures by mixing speech and audio data directly according to Section~\ref{sssec:snr}. Besides shuffling the channel indices, we also shuffle the SNR for each channels at each time step in the time-frequency domain. With such a setup, we aim to model a dynamic environment in which some sensors move around the signal of interest or the signal received by the sensors is disturbed intermittently.
\begin{figure}[htb]
\begin{minipage}[b]{1.0\linewidth}
  \centering
  \centerline{\includegraphics[width=8.5cm]{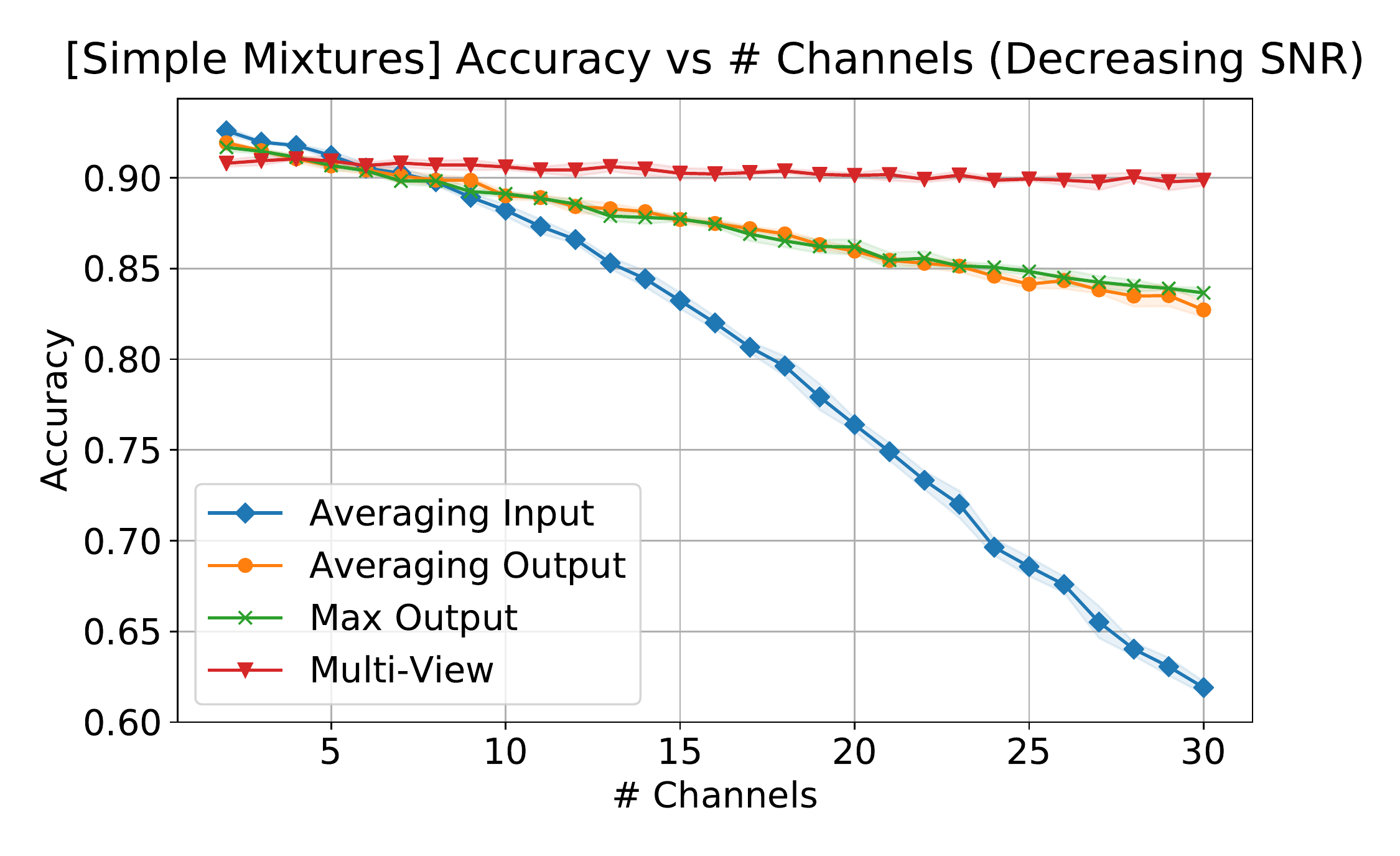}}
  \centerline{(a) Decreasing SNR}\medskip
\end{minipage}
\begin{minipage}[b]{1.0\linewidth}
  \centering
  \centerline{\includegraphics[width=8.5cm]{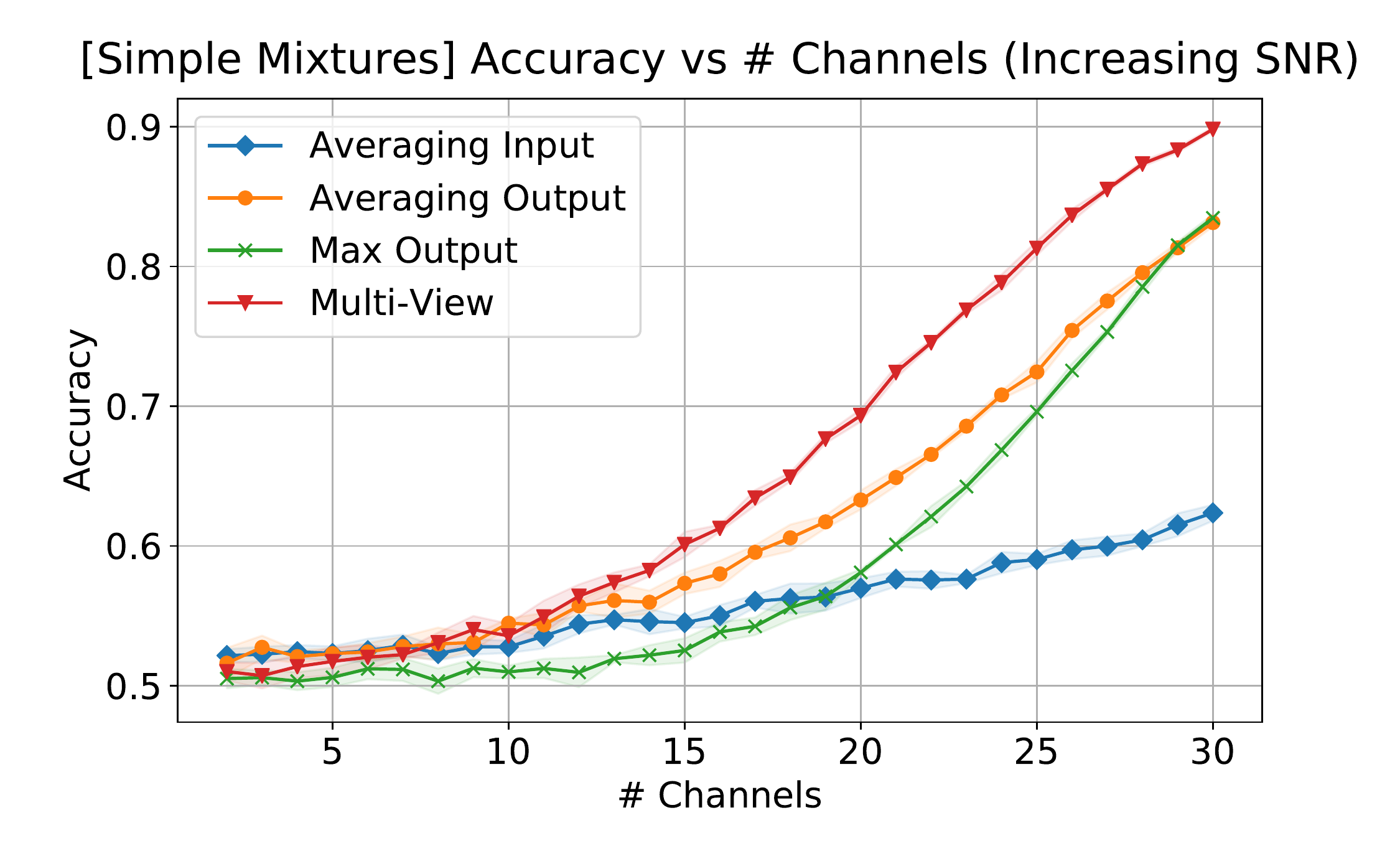}}
  \centerline{(b) Increasing SNR}\medskip
\end{minipage}
%
\caption{Test accuracy of the MVN and the baseline models on audio classification using simple mixtures. The top and bottom plots correspond to the decreasing and increasing SNR scenarios. The x-axis on the plot denotes the number of mixture channels. The y-axis indicates the accuracy of each model's prediction. Each dotted curve shows the mean of the accuracy for five runs of evaluation, and the shaded area represents one standard deviation from the mean.}
\label{fig:exp1}
\end{figure}
Figure~\ref{fig:exp1} shows the prediction accuracy of the MVN and the baseline models from 2 to 30 channels with decreasing SNR. All models are trained on four-channel mixtures with SNRs uniformly spaced between -5 and 5dB. For decreasing SNR, the SNR value decreases by 1 dB for each additional channel. The models have similar performances when the number of channels is less than 10; however, as it goes beyond, the performance of the MVN is more stable than the baseline models. The result shows that the MVN is least affected by the channels with low SNR values compared to the baseline models. For increasing SNR, each extra channel is 1 dB higher than the previous channel. The MVN takes the fewest channels to achieve some desired accuracy, indicating that the MVN collects information more effectively than the baseline models. Moreover, the MVN is able to generalize from training on a fixed number of channels and a limited range of SNRs to testing on a varied number of channels with a large range of SNRs.
\subsubsection{Room Simulation}
\label{sssec:room}
In this experiment, we use pyroomacoustics \cite{scheibler2018pyroomacoustics} to model a $20m$ by $20m$ room with the image source model using fourth order echoes. We train the model on many microphone, speaker, noise source geometries and test on unseen ones. To construct a simulated room we first simulate a noiseless moving speech source. Then, using the same microphone geometry we simulate a noise source randomly placed on a grid. To construct a mixture we take a speech recording and a noise recording which correspond to the same microphone geometry and mix them at some SNR. The separate simulation of noise and speech lets us mix and match to construct training and testing environments without having to explicitly simulate every combination. For training we use mixtures with per channel SNRs between -5 and 5 dB then at test time we experiment with both ``decreasing SNR" and ``increasing SNR" as described in Section~\ref{sssec:snr}. All speakers are modeled as point sources. The simulated room impulse responses where generally five times the length of an STFT window.

\begin{figure}
    \centering
    \includegraphics[width=8.5cm]{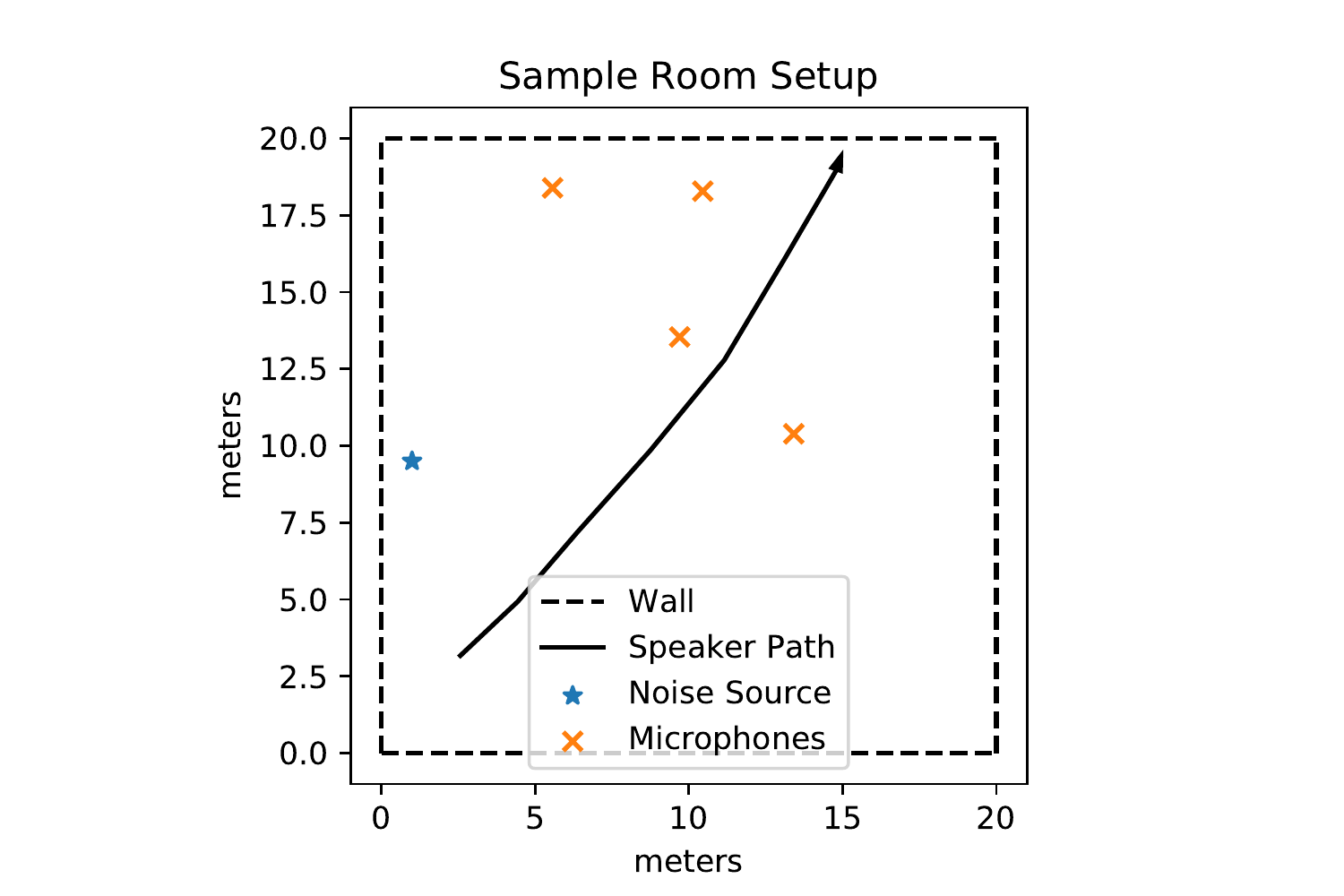}
    \caption{Sample setup with one noise source, four microphones and a moving speech source in a 20 by 20 meter room. Diffuse noise not pictured.}
    \label{fig:room}
\end{figure}

Inside the simulated rooms we place $2$ to $30$ microphones, a noise source and a diffuse noise. The speech source moves in a noisy linear path from one corner of the room to another. The indices of the microphones are randomized. Figure \ref{fig:room} shows one possible configuration of a simulated room.
\begin{figure}[htb]
\begin{minipage}[b]{1.0\linewidth}
  \centering
  \centerline{\includegraphics[width=8.5cm]{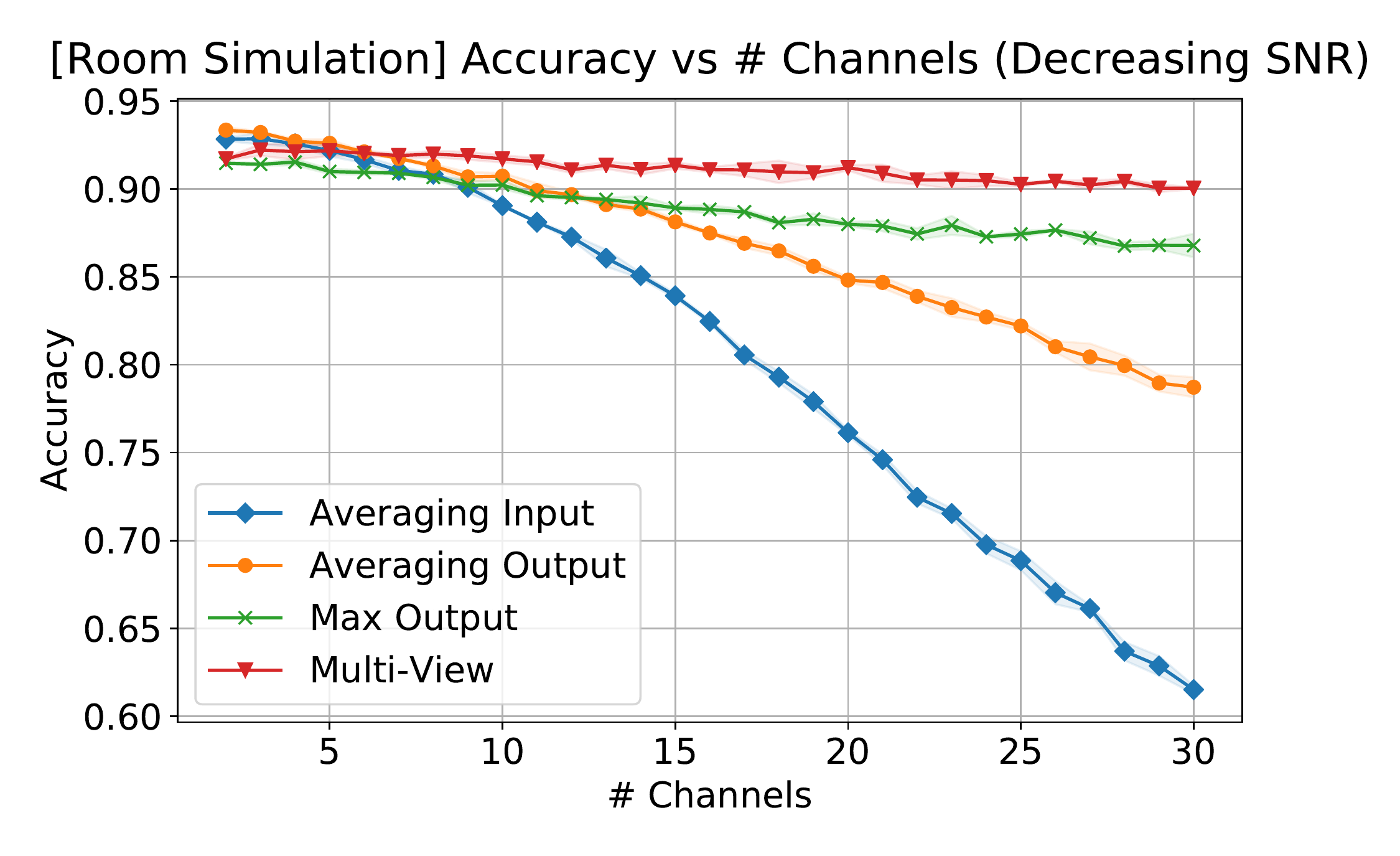}}
  \centerline{(a) Decreasing SNR}\medskip
\end{minipage}
\begin{minipage}[b]{1.0\linewidth}
  \centering
  \centerline{\includegraphics[width=8.5cm]{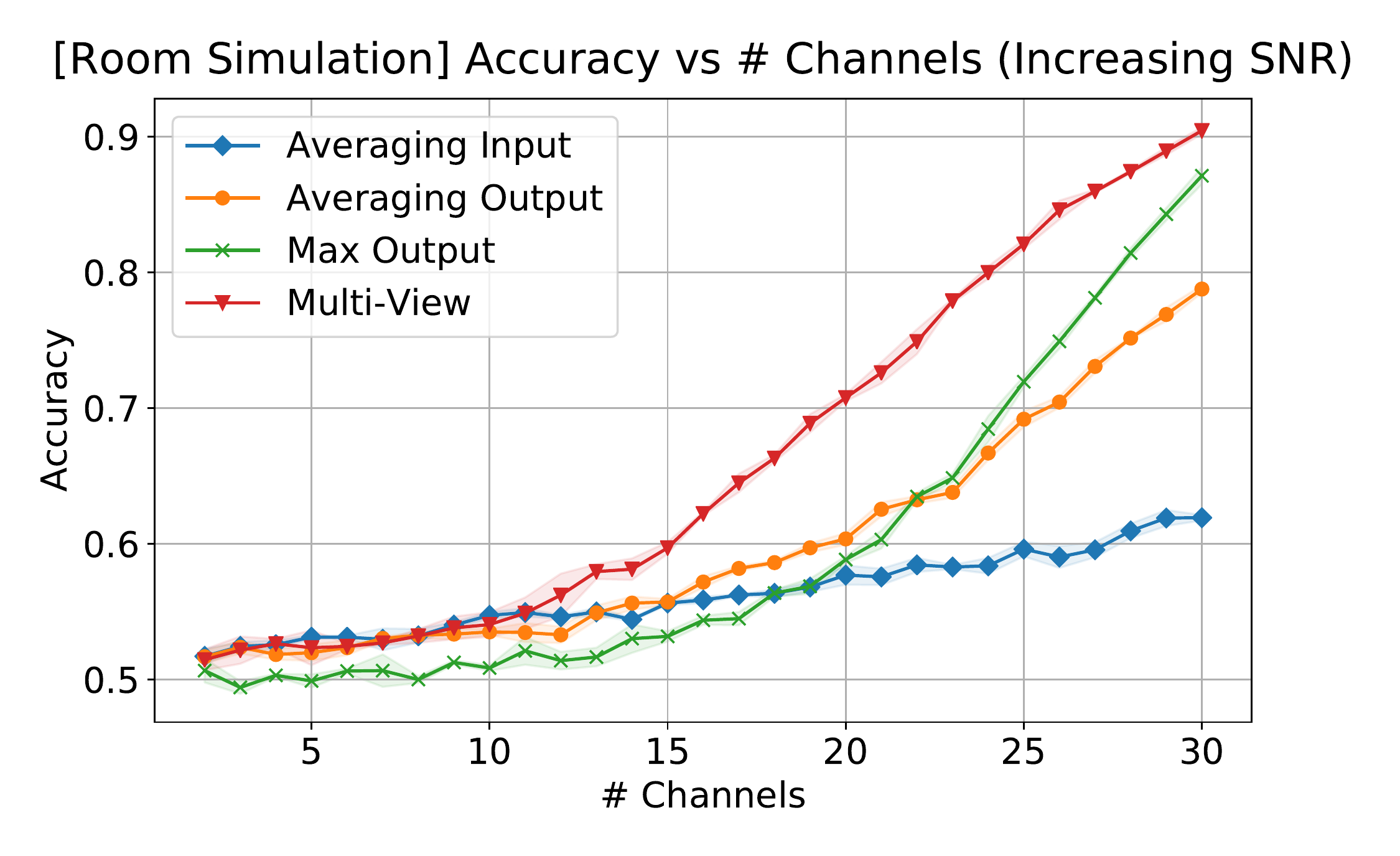}}
  \centerline{(b) Increasing SNR}\medskip
\end{minipage}
\hfill
%
\caption{Test accuracy of the MVN and the baseline models on audio classification using the room simulation. The top and bottom plots correspond to the decreasing and increasing SNR scenarios. Each dotted curve shows the mean accuracy over three runs of evaluation, and the shaded area represents one standard deviation from the mean.}
\label{fig:exp2}
\end{figure}
Figure~\ref{fig:exp2} depicts the accuracy of the MVN and the baseline models for speech classification. The mixtures range from 2 to 30 channels. We observe a pattern similar to the experiment with dry sounds in Section~\ref{sssec:syn}. Among the four models, the MVN is least affected by noise, room impulse responses, and it collects information most effectively from channels that may have high SNRs. 

\section{Conclusion}
\label{sec:conc}
We have proposed a neural network method for multi-channel audio classification using an RNN that unrolls across both channels and time. We demonstrate that the proposed architecture can be deployed to unseen numbers of channels and unseen room geometries at test time. The system is robust to noisy channels in a highly dynamic environment, making it unnecessary to eliminate certain channels as a preprocessing step. Moreover, the system demonstrates the ability to leverage information effectively from a limited number of clean channels when they appear among many noisy channels. As such, this model is capable of being trained once and then deployed in settings with a dynamically changing number of sensors (e.g. in an IoT case), without requiring retraining or any modifications. The proposed framework is not limited to binary classification and can be used for multi-class or multi-label classification, as well with any other type of neural network architecture as long as the channel unrolling takes place. We hope that this will form the basis of powerful future systems that have to operate under uncertainty on the number of input channels, as opposed to resorting to simple averaging of voting schemes which are not as adept in taking the data into account.

\nocite{}
\nocite{*}
\newpage
\bibliographystyle{IEEEbib}
\bibliography{bibliography}

\begin{thebibliography}{10}

\bibitem{hershey2017cnn}
Shawn Hershey, Sourish Chaudhuri, Daniel~PW Ellis, Jort~F Gemmeke, Aren Jansen,
  R~Channing Moore, Manoj Plakal, Devin Platt, Rif~A Saurous, Bryan Seybold,
  et~al.,
\newblock ``Cnn architectures for large-scale audio classification,''
\newblock in {\em Acoustics, Speech and Signal Processing (ICASSP), 2017 IEEE
  International Conference on}. IEEE, 2017, pp. 131--135.

\bibitem{vu2016acoustic}
Toan~H Vu and Jia-Ching Wang,
\newblock ``Acoustic scene and event recognition using recurrent neural
  networks,''
\newblock {\em Detection and Classification of Acoustic Scenes and Events},
  vol. 2016, 2016.

\bibitem{choi2016automatic}
Keunwoo Choi, George Fazekas, and Mark Sandler,
\newblock ``Automatic tagging using deep convolutional neural networks,''
\newblock {\em arXiv preprint arXiv:1606.00298}, 2016.

\bibitem{xu2017convolutional}
Yong Xu, Qiuqiang Kong, Qiang Huang, Wenwu Wang, and Mark~D Plumbley,
\newblock ``Convolutional gated recurrent neural network incorporating spatial
  features for audio tagging,''
\newblock in {\em Neural Networks (IJCNN), 2017 International Joint Conference
  on}. IEEE, 2017, pp. 3461--3466.

\bibitem{Dekkers2017}
Gert Dekkers, Steven Lauwereins, Bart Thoen, Mulu~Weldegebreal Adhana, Henk
  Brouckxon, Toon van Waterschoot, Bart Vanrumste, Marian Verhelst, and Peter
  Karsmakers,
\newblock ``The {SINS} database for detection of daily activities in a home
  environment using an acoustic sensor network,''
\newblock in {\em Proceedings of the Detection and Classification of Acoustic
  Scenes and Events 2017 Workshop (DCASE2017)}, November 2017, pp. 32--36.

\bibitem{inoue2018domestic}
Tadanobu Inoue, Phongtharin Vinayavekhin, Shiqiang Wang, David Wood, Nancy
  Greco, and Ryuki Tachibana,
\newblock ``Domestic activities classification based on cnn using shuffling and
  mixing data augmentation,'' 2018.

\bibitem{Tanabe2018}
Ryo Tanabe, Takashi Endo, Yuki Nikaido, Takeshi Ichige, Phong Nguyen, Yohei
  Kawaguchi, and Koichi Hamada,
\newblock ``Multichannel acoustic scene classification by blind
  dereverberation, blind source separation, data augmentation, and model
  ensembling,''
\newblock Tech. {R}ep., DCASE2018 Challenge, September 2018.

\bibitem{rnn}
Tomas Mikolov, Martin Karafiát, Lukás Burget, Jan Cernocký, and Sanjeev
  Khudanpur,
\newblock ``Recurrent neural network based language model.,''
\newblock in {\em INTERSPEECH}, Takao Kobayashi, Keikichi Hirose, and Satoshi
  Nakamura, Eds. 2010, pp. 1045--1048, ISCA.

\bibitem{parascandolo2016recurrent}
Giambattista Parascandolo, Heikki Huttunen, and Tuomas Virtanen,
\newblock ``Recurrent neural networks for polyphonic sound event detection in
  real life recordings,''
\newblock {\em arXiv preprint arXiv:1604.00861}, 2016.

\bibitem{kim2017acoustic}
Hyoung-Gook Kim and Jin~Young Kim,
\newblock ``Acoustic event detection in multichannel audio using gated
  recurrent neural networks with high-resolution spectral features,''
\newblock {\em ETRI Journal}, vol. 39, no. 6, pp. 832--840, 2017.

\bibitem{li2016neural}
Bo~Li, Tara~N Sainath, Ron~J Weiss, Kevin~W Wilson, and Michiel Bacchiani,
\newblock ``Neural network adaptive beamforming for robust multichannel speech
  recognition.,''
\newblock in {\em INTERSPEECH}, 2016, pp. 1976--1980.

\bibitem{xiao2016deep}
Xiong Xiao, Shinji Watanabe, Hakan Erdogan, Liang Lu, John Hershey, Michael~L
  Seltzer, Guoguo Chen, Yu~Zhang, Michael Mandel, and Dong Yu,
\newblock ``Deep beamforming networks for multi-channel speech recognition,''
\newblock in {\em Acoustics, Speech and Signal Processing (ICASSP), 2016 IEEE
  International Conference on}. IEEE, 2016, pp. 5745--5749.

\bibitem{mvn}
Jonah Casebeer, Brian Luc, and Paris Smaragdis,
\newblock ``Multi-view networks for denoising of arbitrary numbers of
  channels,''
\newblock {\em CoRR}, vol. abs/1806.05296, 2018.

\bibitem{timit}
J.~S. Garofolo, L.~F. Lamel, W.~M. Fisher, J.~G. Fiscus, D.~S. Pallett, and
  N.~L. Dahlgren,
\newblock ``Darpa timit acoustic phonetic continuous speech corpus cdrom,''
  1993.

\bibitem{noise-16k}
Ding Liu, Paris Smaragdis, and Minje Kim,
\newblock ``Experiments on deep learning for speech denoising,''
\newblock {\em Proceedings of the Annual Conference of the International Speech
  Communication Association, INTERSPEECH}, pp. 2685--2689, 1 2014.

\bibitem{adam}
Diederik~P. Kingma and Jimmy Ba,
\newblock ``Adam: {A} method for stochastic optimization,''
\newblock {\em CoRR}, vol. abs/1412.6980, 2014.

\bibitem{scheibler2018pyroomacoustics}
Robin Scheibler, Eric Bezzam, and Ivan Dokmani{\'c},
\newblock ``Pyroomacoustics: A python package for audio room simulation and
  array processing algorithms,''
\newblock in {\em 2018 IEEE International Conference on Acoustics, Speech and
  Signal Processing (ICASSP)}. IEEE, 2018, pp. 351--355.

\bibitem{snr}
D.~H. Johnson,
\newblock ``{S}ignal-to-noise ratio,''
\newblock {\em Scholarpedia}, vol. 1, no. 12, pp. 2088, 2006,
\newblock revision \#91770.

\end{thebibliography}
\end{document}